\documentclass[twocolumn,showpacs,aps,prd,amssymb,nofootinbib,nobibnotes,floatfix,longbibliography]{aastex631}

\usepackage{newtxtext,newtxmath}
\usepackage{ae,aecompl}
\usepackage[mathscr]{euscript}

\usepackage[T1]{fontenc}
\usepackage{hyperref}
\usepackage{amsfonts}
\usepackage{amsmath}
\usepackage{mathrsfs}
\usepackage{epsfig}
\usepackage{graphicx}
\usepackage{url}
\usepackage{hyperref}
\usepackage{float}
\usepackage{color}
\usepackage[utf8]{inputenc} 
\usepackage{epsf}
\usepackage{comment}
\usepackage{slashed}
\usepackage{footmisc}
\usepackage{lipsum}
\definecolor{linkcolor}{rgb}{0.0,0.3,0.5}
\usepackage[caption=false]{subfig}

\usepackage{color}
\definecolor{cerulean}{rgb}{0.0, 0.48, 0.65}

\definecolor{navy}{rgb}{0.2, 0.0, 1.0}

\definecolor{jungle}{rgb}{0.0, 0.5, 0.0}

\usepackage{color}

\definecolor{orange}{rgb}{1,0.5,0}
\definecolor{orangeB}{rgb}{1,0.7,0}

\usepackage[mathlines]{lineno}

\begin{document}

\preprint{APS/123-QED}

\title{Environmental effects in stellar mass gravitational wave sources II:\\
 Enhanced detectability of phase shifts in eccentric sub-populations.}

\author{Lorenz Zwick}
\affiliation{Niels Bohr International Academy, The Niels Bohr Institute, Blegdamsvej 17, DK-2100, Copenhagen, Denmark}
\affiliation{Center of Gravity, Niels Bohr Institute, Blegdamsvej 17, 2100 Copenhagen, Denmark.}
\author{Kai Hendriks}
\affiliation{Niels Bohr International Academy, The Niels Bohr Institute, Blegdamsvej 17, DK-2100, Copenhagen, Denmark}
\affiliation{Center of Gravity, Niels Bohr Institute, Blegdamsvej 17, 2100 Copenhagen, Denmark.}
\author{Pankaj Saini}
\affiliation{Niels Bohr International Academy, The Niels Bohr Institute, Blegdamsvej 17, DK-2100, Copenhagen, Denmark}
\affiliation{Center of Gravity, Niels Bohr Institute, Blegdamsvej 17, 2100 Copenhagen, Denmark.}

\author{János Takátsy}
\affiliation{Institut für Physik und Astronomie, Universität Potsdam, Haus 28, Karl-Liebknecht-Str. 24-25, Potsdam, Germany}

\author{Connar Rowan}
\affiliation{Niels Bohr International Academy, The Niels Bohr Institute, Blegdamsvej 17, DK-2100, Copenhagen, Denmark}

\author{Johan Samsing}
\affiliation{Niels Bohr International Academy, The Niels Bohr Institute, Blegdamsvej 17, DK-2100, Copenhagen, Denmark}
\affiliation{Center of Gravity, Niels Bohr Institute, Blegdamsvej 17, 2100 Copenhagen, Denmark.}

\author{Jakob Stegmann}
\affiliation{Max-Planck-Institut für Astrophysik, Karl-Schwarzschild-Str. 1, 85748 Garching, Germany}

\shorttitle{Eccentric GW populations and Environmental effects}
\shortauthors{Zwick et al.}

\date{\today}

\begin{abstract}
We demonstrate that the properties of eccentric gravitational wave (GW) signals enhance the detectability of GW phase shifts caused by environmental effects (EEs): The signal-to-noise ratio (SNR) of EEs can be boosted by up to $\ell_{\rm max}^{1 - n}$ with respect to corresponding circular signals, where $\ell_{\rm max}$ is the highest modeled eccentric GW harmonic and $n$ is the frequency scaling of the GW dephasing prescription associated to the EE. We investigate the impact on a population level, adopting plausible eccentricity distributions for binary sources observed by LIGO/Virgo/Kagra (A+ and A\# sensitivities), as well as Cosmic Explorer (CE) and the Einstein Telescope (ET). For sources in the high eccentricity tail of a distribution ($e \gtrsim 0.2$ at 10 Hz), phase shifts can systematically be up to $\ell_{\rm max}^{1 - n}$ times smaller than in a corresponding circular signal and still be detectable. For typical EEs, such as Roemer delays and gas drag, this effect amounts to SNR enhancements that range from $10^2$ up to $10^5$. For CE and ET, our analysis shows that EEs will be an ubiquitous feature in the eccentric tail of merging binaries, regardless of the specific details of the formation channel. Additionally, we find that the joint analysis of eccentricity and phase shift is already plausible in current catalogs if a fraction of binaries merge in AGN migration traps.
\end{abstract}


\section{Introduction}
\label{sec:Introduction}
A few candidate eccentric signals have already been identified in the current LIGO-Virgo-KAGRA (LVK) gravitational wave (GW) catalogs \citep{2021ApJ...921L..31R,2024gupte,Morras2025,2025planas}, and it is likely that more will emerge in the recently released O4a observation run. In terms of astrophysics, this is particularly interesting since the presence of eccentricity in gravitational wave (GW) signals is considered a smoking gun signature of dynamical formation channels for compact object (CO) binaries \citep{2006ApJ...640..156G, 2014ApJ...784...71S, 2017ApJ...840L..14S,silsbee2017,antonini2017, Samsing18a, 2018ApJ...855..124S,
2018MNRAS.tmp.2223S, 2018PhRvD..98l3005R, 2019ApJ...881...41L,2019ApJ...871...91Z, 2019PhRvD.100d3010S, 2019arXiv190711231S, 2019PhRvD..99f3003K,2020shaw,2021ApJ...921L..31R,2023oshea,2024dallamico,StegmannKlencki2025}. The argument is that the presence of eccentricity in even a few signals implies the existence of a larger population of binaries from the same formation channel, where the ones that retain measurable eccentricity represent the tail of a distribution \citep[e.g.,][]{Zevin2021ecc,2024stegmann,StegmannKlencki2025}. If these candidate signals are confirmed to be eccentric, their observation implies the detection of a plethora of eccentric sources in future observation runs, in particular if higher sensitivity is achieved at lower GW frequencies where residual eccentricity is more common: Proposed detectors such as the Einstein Telescope can identify eccentricity as small as $10^{-3}$ at $10$ Hz GW frequency~\citep{Saini:2023wdk}. Moreover, eccentricity is not the only relevant signature that can affect signals from stellar mass sources at lower GW frequencies: In \cite{2025zwick}, hereafter PI, we performed an extensive analysis of an even stronger smoking gun signature of binary environments, i.e. the presence of dephasing in the GW signals \citep{1993chakrabarti,1995ryan,2008barausse,2007levin,kocsis,2014barausse,inayoshi2017,2017meiron,2017Bonetti,2019alejandro,2019randall,2020cardoso,DOrazioGWLens:2020, 2022liu,2022xuan,garg2022,2022cole,2022chandramouli,2022sberna,2023zwick,2023Tiede,2024dyson,2021alejandro,2022destounis,2022cardoso,2020caputo,2022zwick,2023aditya,2024zwicknovel,2025Zwick_ecc,2021andrea,2024basu,2024santoro,2025vicente,2025dyson,2025destounis,2025copparoni,2025torres}. Different environmental effects (EE), such as Roemer delays \citep{2017meiron,2019robson,2024samsing,kai22024}, tidal forces \citep{2022chandramouli}, gas drag and torques \citep{Derdzinksi:2021,garg2022,2022speri,2023zwick,2024garg}, leave characteristic imprints in the phase of GW. The prospect of extracting and distinguishing such signatures is plausible for a substantial fraction of high signal-to-noise (SNR) sources in future ground based detectors, and already in LVK for a subset of extreme outliers \citep{2025zwick}.

Both eccentricity and dephasing emerge as a consequence of strong interactions between CO binaries and their environment. Therefore, it is natural to ask whether eccentric GW sources are most likely to also exhibit dephasing, and vice-versa. In other words, whether eccentricity and EE are highly correlated, and the presence of one can be used to inform priors regarding the other. In addition to this, recent work has highlighted how the properties of eccentric GWs, in particular the presence of eccentric harmonics, can be leveraged to substantially increase the possibility of extracting environmental dephasing with respect to a circular GW with the same SNR \citep{2023PhRvD.107d3009X,pedo}, in a manner that is reminiscent of the increased capacity to measure binary vacuum parameters \citep{2020moore}. This aspect has never been thoroughly investigated in the context of realistic populations of merging stellar mass binary sources. Thus, the leading question of this work is posed: Does the existence of eccentric populations of merging CO binaries enhance the prospect of discerning EE in both current and future ground-based GW detectors?

The paper is structured as follows: In section \ref{sec:Methods}, we outline the basic ingredients required for the calculations presented in this work, including eccentric waveforms, dephasing prescriptions and signal-to-noise estimates. We note that this article is a companion piece to PI, where many basic elements are explained in further detail. In section \ref{sec:Det_increases}, we compute the increase in sensitivity to EEs that comes as a result of the interaction of dephasing and eccentric harmonics and quantify the consequences for realistic binary eccentricity distributions. In section \ref{sec:conclusions}, we conclude by commenting on how our findings can inform searches for EEs in current and future GW signal catalogs.

\begin{figure}
    \centering
    \includegraphics[width=1\linewidth]{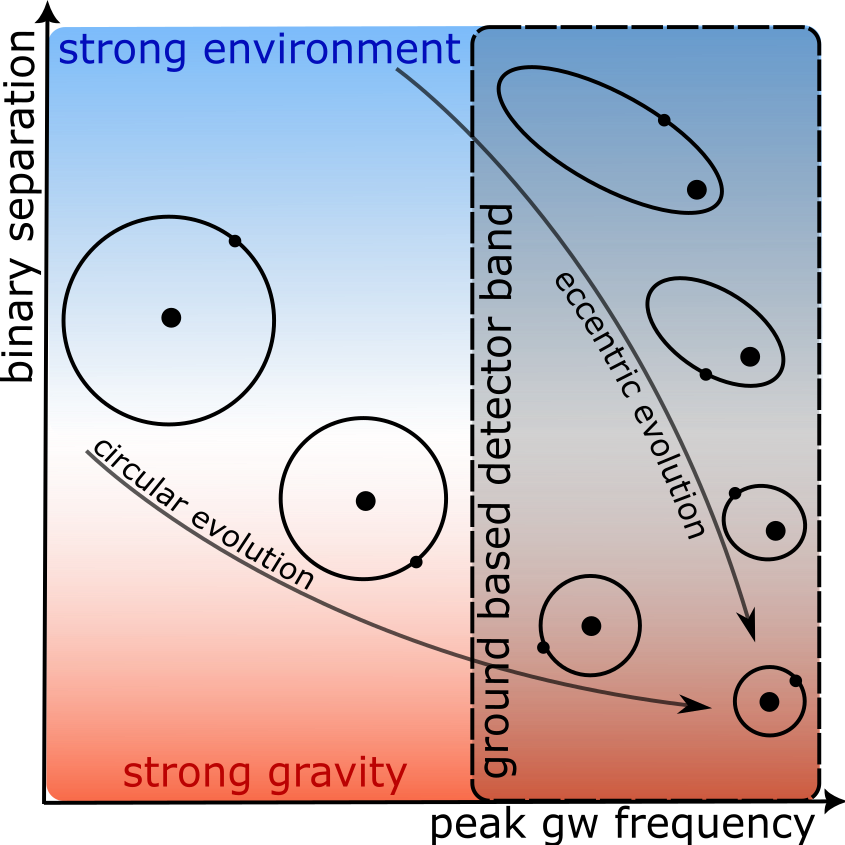}
    \caption{Illustration of the fundamental insight motivating this work. Eccentric binaries radiate GW at the characteristic frequency associated to pericenter passage. Therefore, they produce GW at high frequencies (in particular at the characteristic "peak" frequency) over a larger portion of their entire evolution. They enter the sensitive frequency band of ground based GW detectors at a wider separation, at which environmental effects are strong with respect to relativistic effects. Therefore, we expect a stronger trace of binary environments in the GW signal of eccentric sources.}
    \label{fig:illustration}
\end{figure}

\section{Methodology}
\label{sec:Methods}
\subsection{Eccentric waveforms}

\begin{figure}
    \centering
    \includegraphics[width=1\linewidth]{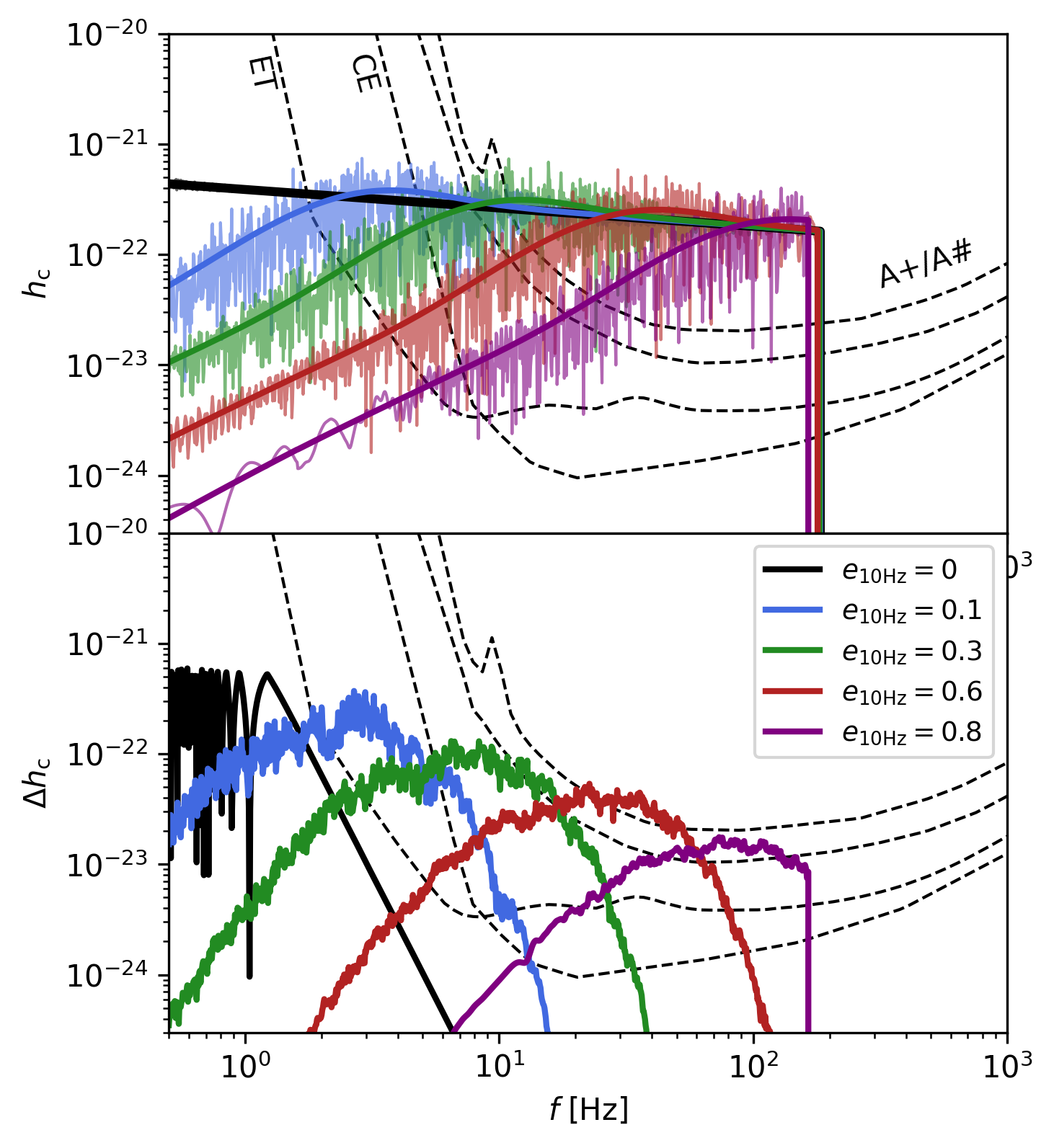}
    \caption{Top panel: Characteristic strain tracks (given by $\tilde{h}(f)\times f$) of a 8 M$_{\odot}$ + 8 M$_{\odot}$ binary at $z=0.5$ compared to various detector sensitivity curves (dashed lines), for some sample eccentricities at 10 Hz (here meaning when the GW $\ell =2$ harmonic reaches 10 Hz). The solid lines represent the GW envelope, while the thin lines show the interference of the various harmonics. The tracks are truncated at the binary innermost stable circular orbit. Bottom panel: residual strain between a the vacuum waveform and the corresponding waveform with a phase shift of $10^{-3}$ radians at 10 Hz and a frequency scaling of $f^{-13/3}$. Note how the residuals are shifted to higher frequencies as the eccentricity increases, entering the band of GW detectors. These particular waveforms are computed with a maximum of 50 harmonics.}
    \label{fig:ecc_waves}
\end{figure}
Relativistic eccentric waveforms appropriate for parameter inference are a matter of active research \citep{2004memmesheimer,2017PhRvD..95b4038H,2022ramos,2025tousif,2025gamboa,2025morras}. For the purposes of this work, it is necessary to adopt a waveform model that: 1) accurately accounts for the distribution of power across the eccentric harmonics, and 2) provides a reference vacuum phase evolution for each harmonic. Additionally, we also require a GW model that is fast to evaluate, such that SNR estimates may be calculated for order $\mathcal{O}(10^6)$ or more choices of source parameters (which is computationally equivalent to $\mathcal{O}(10^8)$ evaluations of circular waveforms, see also PI).

For these reasons, we adapt a Newtonian, eccentric frequency domain GW waveform based on \cite{2018klein} \citep[see e.g.,][for foundational work]{2009yunes,2014yunes}. In the eccentric case, the time-domain GW strain is given as a sum of harmonics:
\begin{equation}
    h(t) = \sum_{\ell=1}^{\infty} \mathcal{A}_\ell(t) \exp\left[ -i\ell \Phi(t) \right] \: ,
    \label{eq:Ecc_waveform}
\end{equation}
where $h(t)$ is the GW strain at the time $t$,  $\Phi$ is a time-domain phase, while $\ell$ denotes the harmonics. The coefficients $\mathcal{A}_\ell$ are given in terms of Bessel functions \citep{1963peters,peters1964,2009yunes}, and determined by the binary eccentricity, and the projection of the elliptic orbit on the observer's frame (beyond the usual parameters such as chirp mass and luminosity distance). The projections can vary due to the precession of the orbit, producing a characteristic splitting in the spectrum of the emitted GW \citep{2018klein}. However, in this work we will instead assume for simplicity that they are constant and that the binary is optimally orientated with respect to the observer. While this enhances the average SNR with respect to randomized orientation angles {by a factor $5/2$ due to the antenna pattern of ground based detectors \citep{2018maggiore}}, we will always compare our results with the corresponding, optimally orientated circular case. Therefore, this simplification only minimally affects our results, which are intended as a comparison between circular and eccentric sources. 

The phase $\Phi$ in Eq.~\eqref{eq:Ecc_waveform} evolves as:
\begin{equation}
    \Phi(t) = \int^t 2\pi F(t') \, \mathrm{d}t' \: ,
\end{equation}
where $F$ is the orbital frequency of the binary. Similarly to the circular case, the stationary phase approximation \citep{1994cutler} can be used to transform Eq.~\eqref{eq:Ecc_waveform} into the frequency domain:
\begin{equation}
    \tilde{h}(f) \approx \sum_{\ell=1}^{\ell_\mathrm{max}} \tilde{\mathcal{A}}_\ell\left( f \right) \exp\left[ i\psi_\ell(f) \right],
\end{equation}
where $f=\ell F$ is the detector-frame GW frequency of the given harmonic. Crucially, this means that the waveform is evaluated at different binary frequencies $F_{\ell} = f/\ell$, corresponding to different epochs of the binary evolution. In other words, multiple earlier stages of the binary lifetime will produce power at high frequencies, due to the presence of the GW harmonics (see also Fig. \ref{fig:illustration}). Note that here we also introduce a maximum harmonic $\ell_{\rm max}$, discussed more thoroughly in section \ref{sec:Methods:maxell}. The phases $\psi_\ell(f)$ are simply multiples of the Fourier phase $\Psi$:
\begin{equation}
    \psi_\ell (f) = \ell \Psi\left( f \right) \: ,
\end{equation}
where:
\begin{equation}
    \Psi(F) =  2\pi F t(F) - \Phi(F) \: ,
\end{equation}
and the time as a function of frequency $t(F)$ and the phase $\Phi(F)$ are given by the following integrals:
\begin{align}
    t(F) &= \int^F \dot{F}^{-1} \, \mathrm{d}F' \: , \label{eq:tF} \\
    \Phi(F) &= 2\pi \int^F F' \dot{F}^{-1} \, \mathrm{d}F' \label{eq:PhiF} \: .
\end{align}
The frequency chirp of the binary $\dot{F}$ is described by the seminal result of \cite{peters1964}:
\begin{align}
    \dot{F} = \dot{F}^{e=0}\times \mathcal{F}(e)
\end{align}
where:
\begin{equation}
    \dot{F}^{e=0}(F) = \frac{96}{5}(2\pi)^{8/3} \left(\frac{G\mathcal{M}}{c^3}\right)^{5/3}F^{11/3} \: ,
    \label{eq:dFvace0}
\end{equation}
and the enhancement function $\mathcal{F}(e)$ reads:
\begin{align}
    \mathcal{F}(e) = \left(1+\frac{73}{24}e^2 + \frac{37}{96}e^4 \right)(1-e^2)^{-7/2} \: .
    \label{eq:Fe}
\end{align}
To evaluate the integrals in Eqs.~\eqref{eq:tF} and \eqref{eq:PhiF} numerically, we require an additional formula for $e(F)$.
To leading order and in the small eccentricity limit $e(F)$ reads:
\begin{equation}
    e(F) \approx e_\mathrm{in} \left( \frac{F}{F_\mathrm{in}} \right)^{-19/18} \: ,
\end{equation}
where $F_\mathrm{in}$ is a reference orbital frequency. Higher order expansions can be found in e.g. \cite{2009yunes}, and a fitting function is also often used. An exact solution for Eqs.~\eqref{eq:tF} and \eqref{eq:PhiF} can also be given in terms of special functions, as detailed in \cite{2018klein}. Here we adopt the latter approach, such that we do not incur in numerical issues when analysing eccentricities of order unity.

We show several illustrative waveforms in the top panel of Fig. \ref{fig:ecc_waves}, for various reference eccentricities at 10 Hz. Note how increasing the eccentricity shifts the power distribution to higher frequencies, though the total emitted power remains roughly constant. The bottom panel of the figure also shows the residual power in the difference between vacuum waveforms and a waveform with dephasing due to EE, as described in section \ref{sec:Methods:dephasing}.

\subsection{SNR calculations for eccentric sources}
The SNR of a Fourier domain waveform is a measure of the total power emitted in GW compared to the sensitivity of a given detector. It is defined as \citep{2015moore,2018maggiore,2019robson}:
\begin{equation}
    \mathrm{SNR}^2 = 4\int \frac{|\tilde{h}(f)|^2}{S_n(f)} \mathrm{d}f \: ,
\end{equation}
where $S_n(f)$ is the noise power spectral density. In this work, we will consider the projected sensitivity curves of four different near future ground-based detector configurations, i.e LVK A+ \citep{2022ligopl,2025ligopl} and A\# \citep{2024ligosharp}, Cosmic Explorer \citep{2023arXiv230613745E} and the Einstein Telescope \citep{2020maggioreet}. Out of all these configurations, the most interesting difference is the increase in sensitivity of ET at low frequencies, which in general greatly enhances the prospects of detecting EE. As in PI, we will estimate the detectability of EE via the $\delta$SNR criterion \citep[see also][among many other]{kocsis,2023zwick}:
\begin{equation}
    \delta \mathrm{SNR}^2 = 4 \int \frac{|\delta\tilde{h}(f)|^2}{S_n(f)} \mathrm{d}f ,
\end{equation}
where the $\delta \mathrm{SNR}^2$ is required to be larger than a given threshold $\mathcal{C}$. The difference between waveforms is:
\begin{align}
    \delta \tilde{h}(f) &= \tilde{h}_\mathrm{tot}(f) - \tilde{h}_\mathrm{vac}(f),
\end{align}
where $\tilde{h}_\mathrm{tot}(f)$ represents the waveform with EE and $\tilde{h}_\mathrm{vac}(f)$ is the corresponding vacuum waveform. In this work, we will choose a representative value of $\mathcal{C}=3$ to represent a signal with significant dephasing. Note that the advantages and limitations of this criterion and choice of threshold are amply discussed
in PI \citep[see also e.g.,][]{2022speri,2023NatAs...7..943C,owen23,2023zwick,2025thompson}. Most crucially, the criterion does not account for degeneracies, and only serves as an indication regarding the possibility of extracting an EE signature with full parameter inference \footnote{For the purposes of this work, adopting full-parameter inference techniques would be computationally unfeasible, and would not provide any additional clarity.}.For eccentric sources, the $\delta$SNR criterion simply becomes a sum of the difference in the waveform for each harmonic:
\begin{equation}
    \delta\mathrm{SNR}^2 = 4\int \frac{1}{S_n(f)} \Big|\sum_{\ell}\delta \tilde{h}_\ell(f) \Big|^2 \mathrm{d}f ,
    \label{eq:SNR_ecc}
\end{equation}
otherwise remaining identical in nature.

{We note here that for weak EE, the $\delta$SNR of a dephased waveform roughly scales as SNR$\times\delta \psi$, where $\delta \psi$ is the phase shift associated to the EE. In other words, the $\delta$SNR increases linearly with the amplitude of the dephasing. This trend is only broken when the dephasing reaches values of $\sim \pi$, after which the entire SNR available in the signal is saturated. The same consideration applies to each individual harmonic of an eccentric signal, for which the corresponding SNR can saturate whenever the dephasing in each harmonic reaches values of $\sim \pi$. This consideration will be crucial to understand the results presented in section \ref{sec:Det_increases}, and also implies that our results can therefore be scaled linearly to different choices of $\mathcal{C}$. Note also that previous work directly comparing dephasing $\delta$SNR and Bayesian parameter estimation for stellar mass compact objects found that a threshold of 3 roughly corresponds to excluding the non-detection of EE at 90\% confidence \citep{zwicklensing}}.

\subsection{Dephasing prescriptions for environmental effects}
We will first analyse a general family of dephasing prescriptions before treating more specific examples of EE. We adopt the dephasing Family I identified in PI, since it entails the vast majority of common EEs \citep[see also][]{2014barausse,2020cardoso}. In the time domain (denoted by $\delta \phi$), the phase shift reads:
\begin{align}
\label{eq:dephfamI}
    \delta \phi^{I}_{i,j,k;n}(F) &=  A_2 \times \mathcal{M}^{i} \mu^{j} F^{n} \mathcal{F}_k^{I}(e),
\end{align}
where:
\begin{align}
    \mathcal{F}_k^{I}(e) &= e^k\left(\frac{T_{\rm in}}{T_{\rm in}^{e=0}} \right) ^2.
\end{align}
Here $\mathcal{M}$ is the chirp mass of the binary, $\mu$ the reduced mass, $A_{2}$ a coefficient denoting the size of the dephasing at a specific frequency and $T_{\rm in}$ is the binary inspiral timescale at the orbital frequency $F$. The subscript to the coefficient $A_2$ denotes the fact that this refers to the dephasing in the main GW 2-2 mode, which is most often left implicit when referring to circular waveforms.

As special cases to this general prescription, we will consider the most promising EE in terms of detectability which were identified in PI. These consist of Roemer delays as well as various flavors of gas effect. Lifting directly from PI and replacing $f \to 2F$ we write down the dephasing prescription for Roemer delays:
\begin{align}
    \label{eq:Romerdeph}
    \delta\phi_{\rm R} &= \left(\frac{5}{256 \pi^{13/6}}\right)^2\frac{c^9 m_3 (2F)^{-13/3}}{R^2G^{7/3}\mathcal{M}^{10/3}}\mathcal{F}_0^{I}(e),
\end{align}
for Bondi-Hoyle-Littleton (BHL) drag:
\begin{align}
    \delta \phi_{\rm BHL} &= \frac{75}{16384 \pi ^{11/3} }\frac{ c^{10} \mathcal{M}^{5/6} \rho (2F)^{-14/3}}{c_{\rm s}^2 
   G^{5/3} \mu ^{7/2}} \mathcal{F}_0^{I}(e),
\end{align}
and for circumbinary disc (CBD) viscous torques:
\begin{align}
    \delta \phi_{\rm visc} &=- \frac{225 f_{\rm CBD} }{8192 \pi ^{10/3}}\frac{\alpha  c^{10} c_{\rm s}^2\Sigma (2F)^{-16/3}}{
   G^{10/3}\mathcal{M}^{10/3} \mu  },
\end{align}
where the derivation and the intuitions behind these formulas are detailed in PI. Here $m_3$ and $R$ represent a third body mass and distance, while $\rho$, $\Sigma$ and $c_{\rm s}$ represent gas density, surface density and speed of sound. We set the fudge factor $f_{\rm CBD}=1$ and generally assume a viscosity of $\alpha = 0.1$, unless stated otherwise. While the specific details will not be used in this work, the combinations of physical parameters for these prescriptions are crucial to understand the consequences of a detectability increase. They are denoted by the variables $\xi_i$ and are summarised in Table \ref{tab:placeholder}.

\begin{table}[]
    \centering
    \begin{tabular}{c|c}
        EE & $\xi_i$ \\
        \hline
        Roemer & $m_3/R_3^2$ \\
        CBD Torques & $c_{\rm s}^2 \Sigma$ \\
        BHL Drag& $\rho/c_{\rm s}^2$
    \end{tabular}
    \caption{Physical parameters determining the magnitude of various dephasing prescriptions for EE.}
    \label{tab:placeholder}
\end{table}

In this work, as in PI, we will make the simplification that the dephasing prescription here reported map identically to the corresponding Fourier domain dephasing $\delta \psi$. This is exactly correct in the case of Roemer delays, and appropriate up to a prefactor of order unity for dephasing prescriptions that are polynomial in $F$ \citep[see][for a derivation and further clarification]{pedo}. In all cases, the scaling of the dephasing prescription with physical parameters remains unaffected. Finally, we note that in these formulae the orbital frequency $F$ must be appropriately redshifted to account for cosmological distances. This is achieved by replacing $F \to (1 +z)F$, where now $F$ is the observer frame binary orbital frequency.

\subsection{Dephasing in eccentric harmonics}
\label{sec:Methods:dephasing}
In order to appropriately apply a dephasing prescription to an eccentric harmonic in the Fourier domain ( denoted by $\delta \psi$), one must consider two separate effects. The first and least impactful, is that the prescription will be multiplied by the harmonic number $\ell$, just as the total GW phase:
\begin{align}
    \delta \psi_{\ell}(f) = \frac{\ell}{2}\delta \psi(F_{\ell}),
\end{align}
where we normalise to the $\ell=2$ harmonic. Intuitively, this can be understood as relating to the fact that eccentric waveforms present a bursty strain time-series. In a GW with sharper features with respect to the corresponding circular case, the same amount of time delay ($h(t) \to h'(t)=h(t +\delta t)$) will produce a more drastic separation between strain peaks and troughs (in other words, a larger phase shift).

The second and more impactful aspect of dephasing in eccentric harmonics is related to the value of the frequency $F_{\ell}$.  As mentioned previously, many different epochs of the binary's evolution are contributing to the observed power at the detector frequency $f$. This is because each GW harmonic emits power at the detector frequency $f$ whenever the binary is orbiting at $F_\ell = f/\ell$. In particular, power from higher harmonics is emitted further back in time, when the binary is orbiting at large separations where EE have a large effect (see Fig. \ref{fig:illustration}). This aspect is reflected in the dephasing, which we now explicitly write as:
\begin{align}
    \label{eq:dephasingboos}
    &\delta \psi_{\ell}(f) = \frac{\ell}{2}\delta \psi_2(\frac{2f}{\ell}).
\end{align}

Evaluating Eq. \ref{eq:dephasingboos} will have strong consequences on the magnitude of the dephasing in various harmonics: Typical EEs scale strongly with negative powers of the frequency (often referred to as ``negative" PN orders). As an example, consider the scaling of dephasing due to Roemer delays, where in the circular case:
\begin{align}
    \delta \psi_{\rm R ;\, 2}^{e=0} (f) \propto f^{-13/3},
\end{align}
In the eccentric case, we will instead have:
\begin{align}
\label{eq:ellscaling}
    \delta \psi_{{\rm R}; \,\ell} \propto {\ell} \left(\frac{ f}{\ell}\right)^{-13/3},
\end{align}
where we used the fact that $f = \ell F$. Crucially, Eq. \ref{eq:ellscaling} scales approximately as $\ell^{5}$. In more generality, the scaling is:
\begin{align}
    \delta \psi_{\ell} \propto \delta \psi_{2} \times \ell^{1 -n},
\end{align}
showing how, as long as EE are present at all, their effect will be massively boosted in the phase of higher harmonics. Depending on the magnitude of the residual eccentricity of the binary when it enters the detector band, such drastic increases in dephasing can lead to corresponding large increases in the $\delta$SNR of the EE.

\subsection{The choice of a maximum harmonic}
\label{sec:Methods:maxell}
As seen in Eq. \ref{eq:ellscaling}, the enhancement of dephasing due to EE in eccentric signals depends critically on the maximum harmonic $\ell_{\rm max}$ included in the waveform model. This parameter sets the upper limit on the harmonic content used to reconstruct the signal. Physically, increasing $\ell_{\rm max}$ corresponds to resolving the rapid modulations of the GW signal generated near pericenter. Conversely, this corresponds to including a larger portion of the binary lifetime in the GW signal at a given frequency. 

The choice of $\ell_{\rm max}$ is governed by two main factors. First, waveform availability and modeling accuracy: Including higher harmonics requires accurate modeling of the corresponding multipolar structure of the source and reliable numerical or analytical waveform templates. Current relativistic eccentric waveform models are typically truncated at moderate $\ell_{\rm max} \sim10$, both to limit computational cost and because the accuracy of higher-order harmonics becomes uncertain beyond this range \citep{2017PhRvD..95b4038H,moore2019}. Second, observational duration: extracting power from high harmonics effectively requires observing multiple burst-like emissions as the binary passes through pericenter, where each harmonic corresponds to GW emission at a frequency approximately $\ell F$. To coherently accumulate SNR across many harmonics, the detector must observe a sufficiently long portion of the inspiral where these pericenter bursts occur. We can estimate the required duration of an observation required to access a given harmonic $\ell$ as follows. Consider the typical duration of a circular LVK signal at a GW frequency of 10Hz of $\sim 100$ s or less. The inspiral timescale scales with the binary orbital frequency as $F^{-8/3}$. Therefore, the inspiral timescale when the $\ell$-th harmonic reaches 10 Hz is approximately 100 s $\times(2/\ell)^{-8/3}$ without accounting for any reduction due to eccentricity.

In this work, we adopt two representative choices for $\ell_{\rm max}$. The first choice is $\ell_{\rm max} =10$, which corresponds to what is currently achieved in commonly available eccentric waveform templates appropriate for parameter inference \citep[][]{huerta2018,moore2019,2022knee,2022ramos,2025gamboa}. Our claims regarding the detectability of EE will be entirely based on this more conservative value. As a more speculative choice, we also consider $\ell_{\rm max}=50$, imagining a scenario in which burst timing waveform models \citep{2014tai,2017loutrel, 2023isobelr, 2025pankaj} can be precisely matched to current templates while also preserving phase coherency \citep[see][for a novel approach to this problem]{2025tousif_harm}. Using the scaling derived above, we see that these correspond to a maximum required signal duration of at most $\sim$hours and $\sim$weeks, respectively, without accounting for any shortening due to eccentricity. Therefore, we do not expect the accumulation of signal power to be limited by the duration of the experiment. Finally, we note that a choice of $\ell_{\rm max}=10$ roughly corresponds to being able to fully appropriately time domain waveforms for binaries with an eccentricity of $\lesssim 0.7$ according to \citep{2018mooreecc}.
\section{Detectability of dephasing as a function of eccentricity}
\label{sec:Det_increases}
\subsection{Impact on individual GW signals}
\begin{figure*}
    \centering
    \includegraphics[width=1\linewidth]{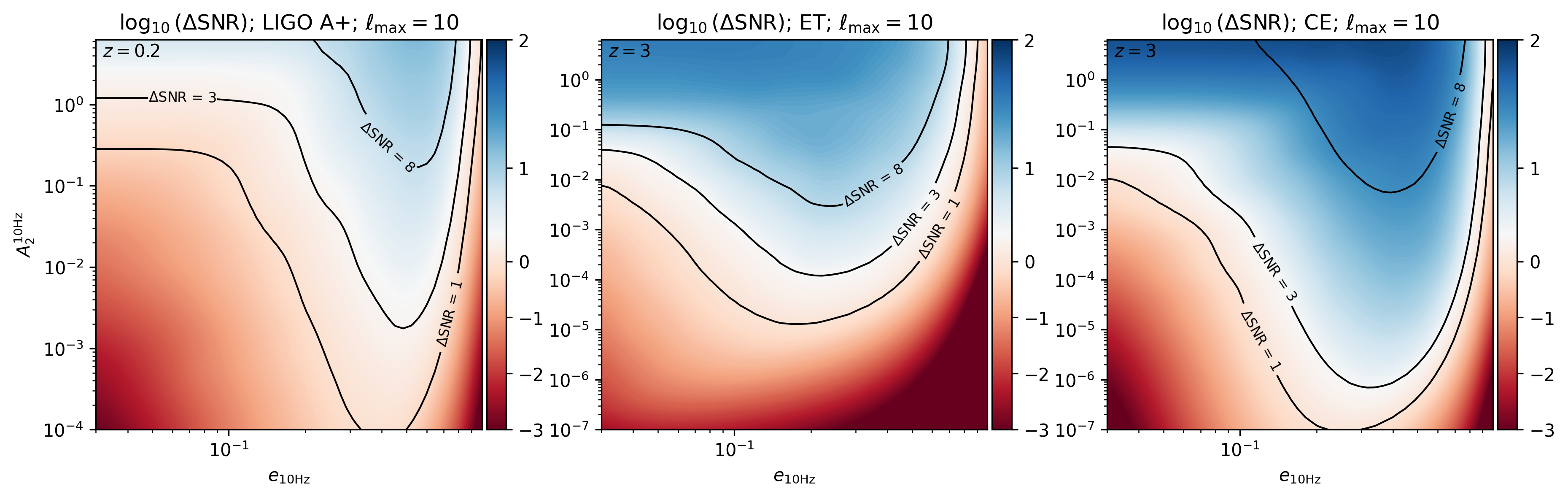}
    \includegraphics[width=1\linewidth]{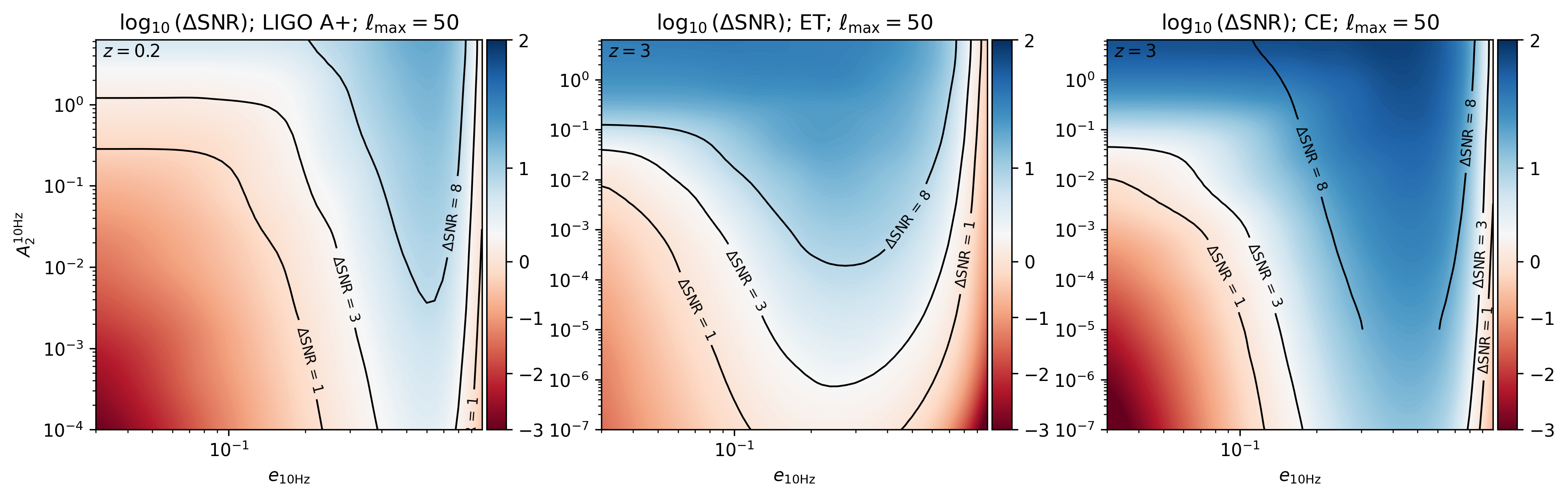}
    \caption{Contour plots for the $\delta$SNR for a binary source of GW with $m_1 = m_2 =$ 8 M$_{\odot}$ located at a typical redshift for the given detector configuration. The contours are computed for a dephasing prescription with $n=-13/3$, and show the dependance of the $\delta$SNR on the magnitude of the dephasing and the eccentricity at 10 Hz. The top row is computed for $\ell_{\rm max}=10$, while the bottom row is for $\ell_{\rm max}=50$. Note how the detectability of the dephasing is greatly increased as soon as significant power is distributed in the higher harmonics of the GW emission, i.e. for eccentricities of $\gtrsim 0.1$. The phase space regions with approximately constant $\delta$SNR as a function of $A_2^{10\rm Hz}$ result from the saturation of the eccentric harmonics (see text).}
    \label{fig:deltaSNR_contour}
\end{figure*}

\begin{figure}
    \centering
    \includegraphics[width=1\linewidth]{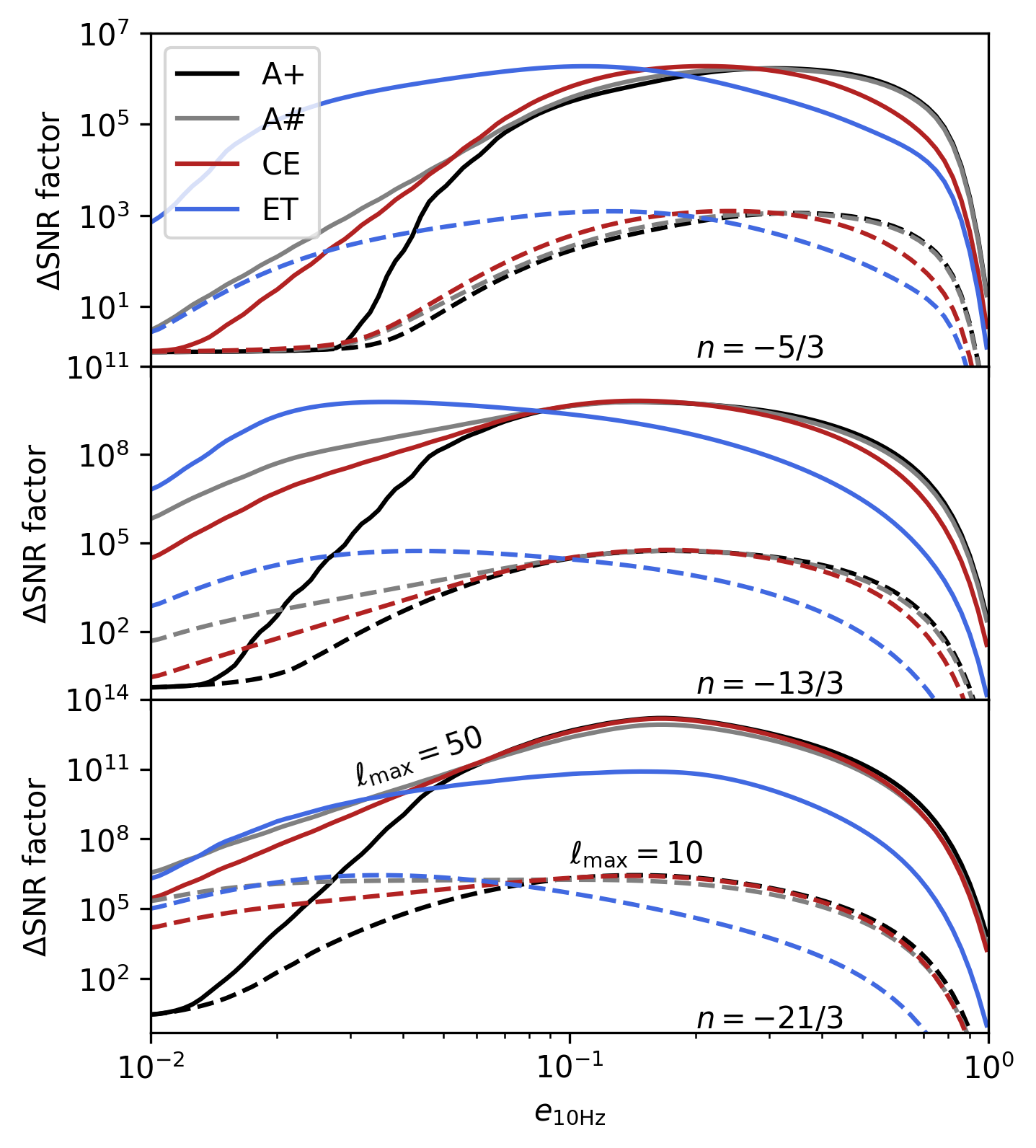}
    \caption{Increases in the $\delta$SNR as a function of the residual eccentricity at 10 Hz with respect to a circular signal. The results are computed for different detectors (coloured lines), dephasing power laws (panels, see Eq. \ref{eq:dephfamI}) and for two representative choices for $\ell_{\rm max}$ (solid and dashed lines). The curves are computed for a dephasing amplitude of $A_2^{10 \rm Hz} =10^{-15}$, ensuring that no eccentric harmonic is ever saturated. Note that the maximum achieved $\delta$SNR factors scales roughly as $\ell_{\rm max}^{1 -n}$, and that these values are reached for a range of moderate eccentricities. {The n values have been chosen to represent the leading scaling of the GW phase ($n=-5/3$), a typical EE ($n=-13/3$) and a very steeply decaying EE ($n=-21/3$) to illustrate the extreme scaling of the dephasing in each harmonic. Indeed,} the $n=-5/3$ results are essentially showing an increased capacity to determine the binary's chirp mass (recall the $F^{-5/3}$ scaling of the vacuum GW phase), a phenomenon already studied in \cite{2020moore}.}
    \label{fig:delta_SNR_n}
\end{figure}
Our aim is to quantify how the properties of dephasing in eccentric harmonics result in an increase in the detectability of EE. More precisely, we investigate how the $\delta$SNR of a dephasing prescription with a given amplitude in $\ell=2$ harmonic can increase by very large factors when the binary retains moderate eccentricity at 10 Hz. For our numerical tests, we choose a representative binary source of GW with $m_1 = m_2= 8$ M$_{\odot}$, located at a redshift of $z=0.2$ for LVK and $z=3$ for CE/ET, respectively (though note that the exact choices do not matter here in terms of our conclusions). We consider a dephasing prescription of the form shown in Eq. \ref{eq:dephfamI}. The strength of the EE is fixed such that the dephasing has a certain value regardless of redshift, when the binary orbital frequency reaches 5 Hz and the $\ell=2$ harmonic reaches 10 Hz in the detector frame:
\begin{align}
    \delta \psi_{\ell=2} = A^{10 {\rm Hz}}_{2} \left(\frac{2F}{10\, {\rm Hz}}\right)^{-n}.
\end{align}
We evaluate $\delta$SNRs according to Eq. \ref{fig:delta_SNR_n} using the waveforms detailed in section \ref{eq:Ecc_waveform}, as a function of the various dephasing parameters and binary eccentricity. We express the results both in terms of the $\delta$SNR directly, as well as the ratio between the $\delta$SNR for eccentric binaries with respect to a corresponding circular binary. We refer to the latter as the $\delta$SNR factor.

In Fig \ref{fig:deltaSNR_contour}, we show the $\delta$SNR results for $n=-13/3$ as a contour plot, for a grid of binary eccentricities at 10 Hz and dephasing amplitudes at 10 Hz. The range of amplitudes goes from 2$\pi$, representing a strong EE that entirely saturates the SNR of the $\ell = 2$ harmonic, to very small numbers representing the presence of very weak EE. For low eccentricities ($\lesssim0.1$), we recover the standard result that EEs cause significant residual signal power whenever the dephasing they induce is of the order few $ \pi$/SNR \citep[see e.g.][]{2023zwick}. For moderate eccentricities of $\gtrsim 0.1$, dephasing amplitudes that are many orders of magnitude smaller can lead to significant $\delta$SNR. This strongly implies that the detectability of the corresponding EE would be greatly enhanced. Interestingly, the plots showcase regions of phase space where the $\delta$SNR remains high almost regardless of the value of $A_2^{10 \rm Hz}$, implying that the results are dominated by the saturation of the eccentric GW harmonics rather than the actual strength of the EE. The results are shown for the two representative choices of $\ell_{\rm max}=10$ and $\ell_{\rm max}=50$. They are qualitatively similar, with the latter showing more drastic effects, as expected.

The transition at $e_{10 \rm Hz} \sim 0.1$ coincides exactly with the fact that higher GW harmonics begin to dominate in terms of emitted power for binaries with eccentricities larger than $\sim 0.1$ \citep{peters1964,2009yunes}. In this regime, the drastic increases in dephasing discussed in section \ref{sec:Methods:dephasing} can result in an entirely saturated SNR of higher GW harmonics. The decrease in $\delta$SNR for extreme eccentricities has instead two origins: The overall diminishment of the SNR of the waveform as well as the suppression in the dephasing for high eccentricities, which scale as the function $\mathcal{F}_k^{I}(e)$ described in section \ref{sec:Methods:dephasing}.

Fig. \ref{fig:delta_SNR_n} presents some similar results, in which we vary the scaling of the dephasing prescription with the frequency, i.e. the parameter $n$. Stronger scaling with the frequency exacerbates the effect of dephasing in higher GW harmonics, thus shifting the maximum $\delta$SNR factor to higher eccentricity and to higher values. Here, the results for the $\delta$SNR factor are computed for an extremely weak EE, with an amplitude $A_2^{10 \rm Hz}=10^{-15}$. This is chosen to ensure that no harmonic is ever saturated, and that the full extent of the increased dephasing in higher harmonics is modeled. Indeed, we observe the strong scaling of the dephasing with the harmonic number $\ell$ by looking at the maximum values reached by the $\delta$SNR factor:
\begin{align}
    \frac{\max\left(\delta{\text {SNR}}(e_{10 \rm Hz})\right)}{\delta{\text {SNR}}(0)} \approx \ell_{\rm max}^{1-n},
\end{align}
and that values of this order are reached for a wide range of eccentricities $0.1 \lesssim e_{10 \rm Hz} \lesssim 0.4$. This is in fact the expected scaling from Eq. \ref{eq:ellscaling}, in the limit where no GW harmonic is saturated. Interestingly, we also observe how different detector configurations affect the results. Due to the wider sensitivity band, ET can benefit from increases in the $\delta$SNR of EE already at smaller reference eccentricities at 10 Hz, and especially for shallower frequency scalings. However, ET looses out on the ability to extracting power from very high harmonics, due to the flatter sensitivity band with respect to the characteristic LVK/CE bucket shape (see Fig. \ref{fig:ecc_waves}).

Overall, our findings indicate how moderately eccentric GW sources will be extremely powerful probes of EE, allowing to discern environmental perturbations several order of magnitudes smaller than what is expected for circular signals. These results are analogous to the increase in the capacity to measure standard vacuum parameters demonstrated in \cite{2020moore} for eccentric vacuum signals, though the results are exacerbated even more due to the stronger scaling of EEs with frequency.

\subsection{Impact on eccentric binary sub-populations}
\label{sec:Det_increases_pop}

\begin{figure*}
    \centering
    \includegraphics[width=0.49\linewidth]{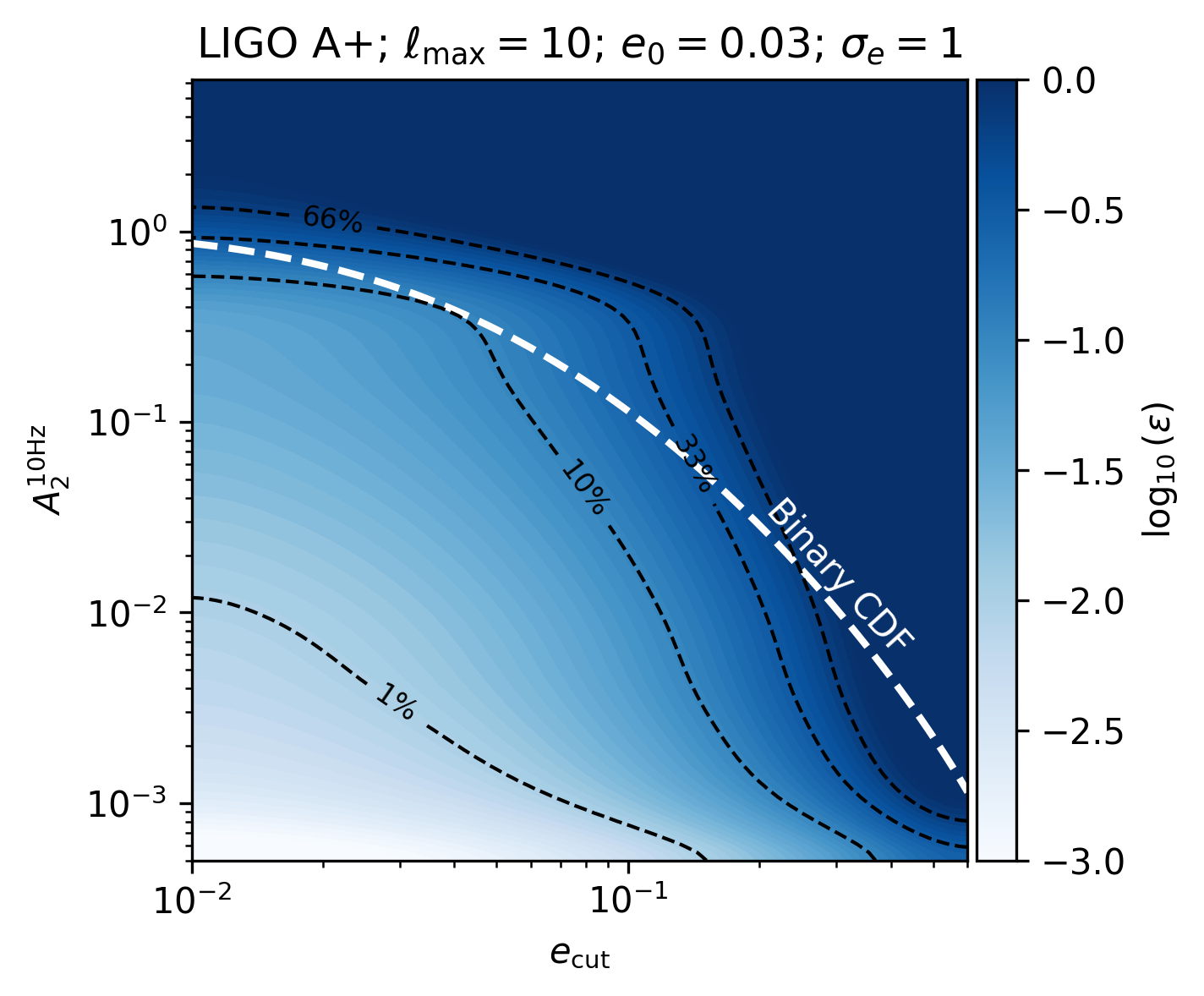} \includegraphics[width=0.49\linewidth]{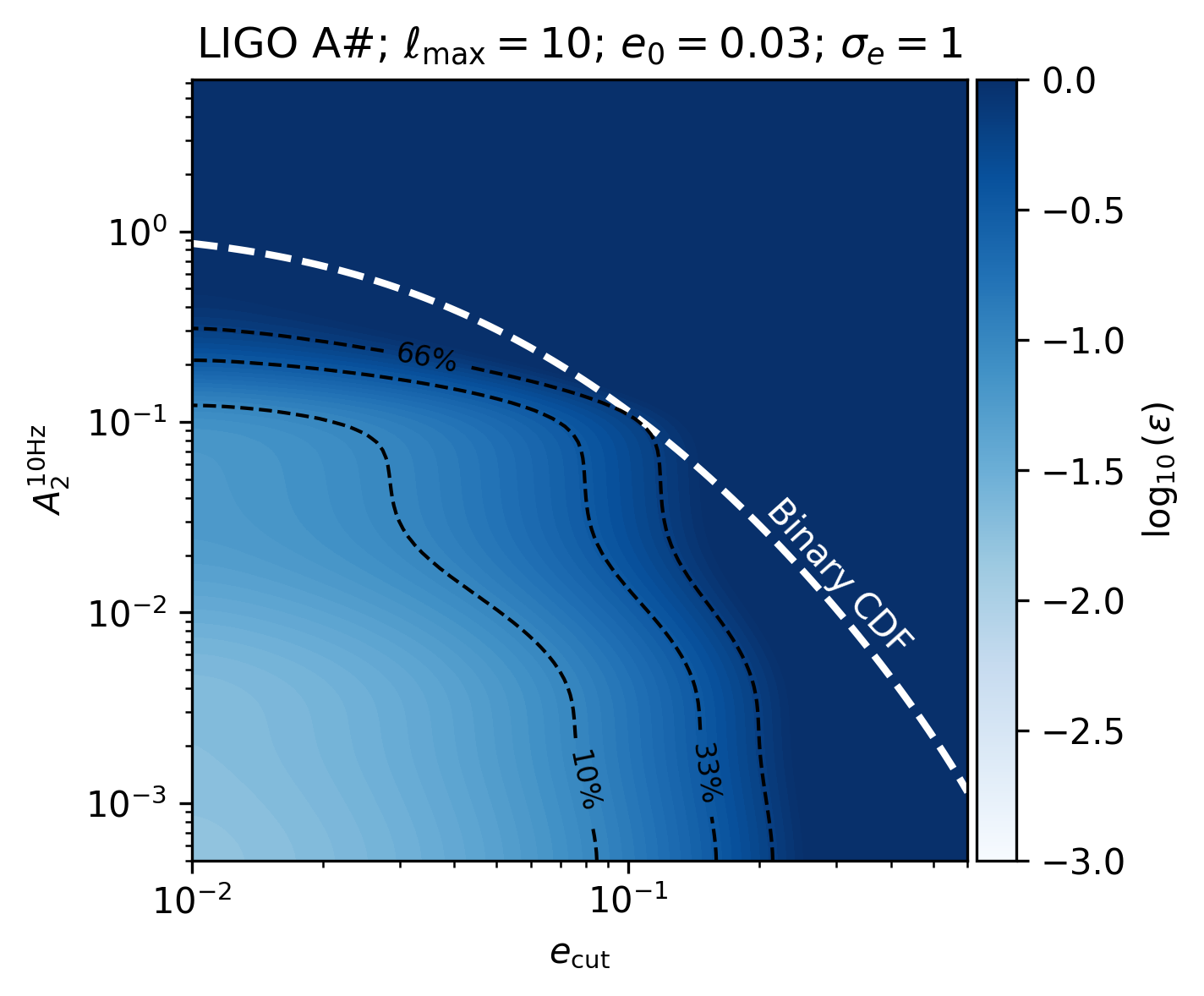}
\includegraphics[width=0.49\linewidth]{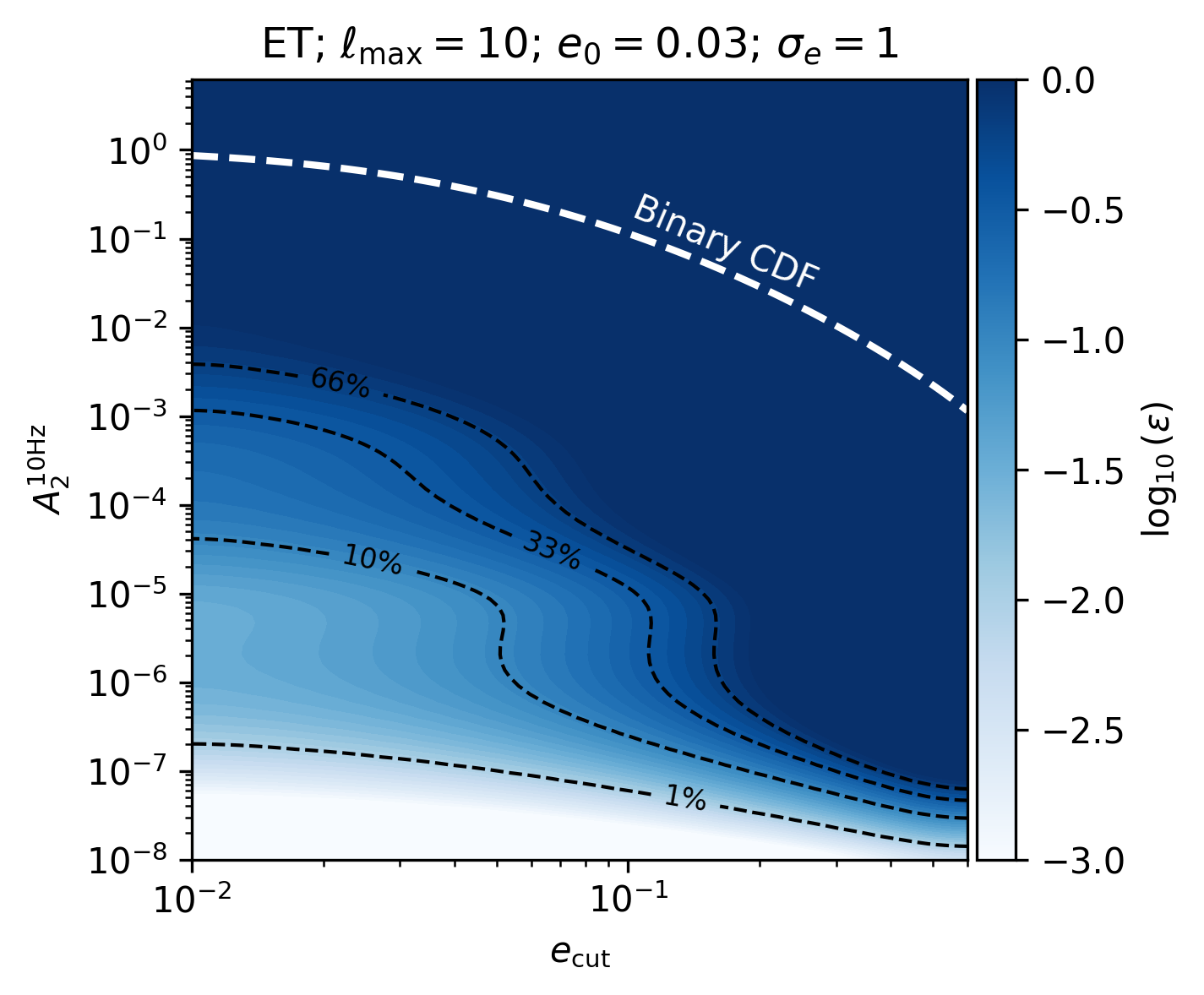} \includegraphics[width=0.49\linewidth]{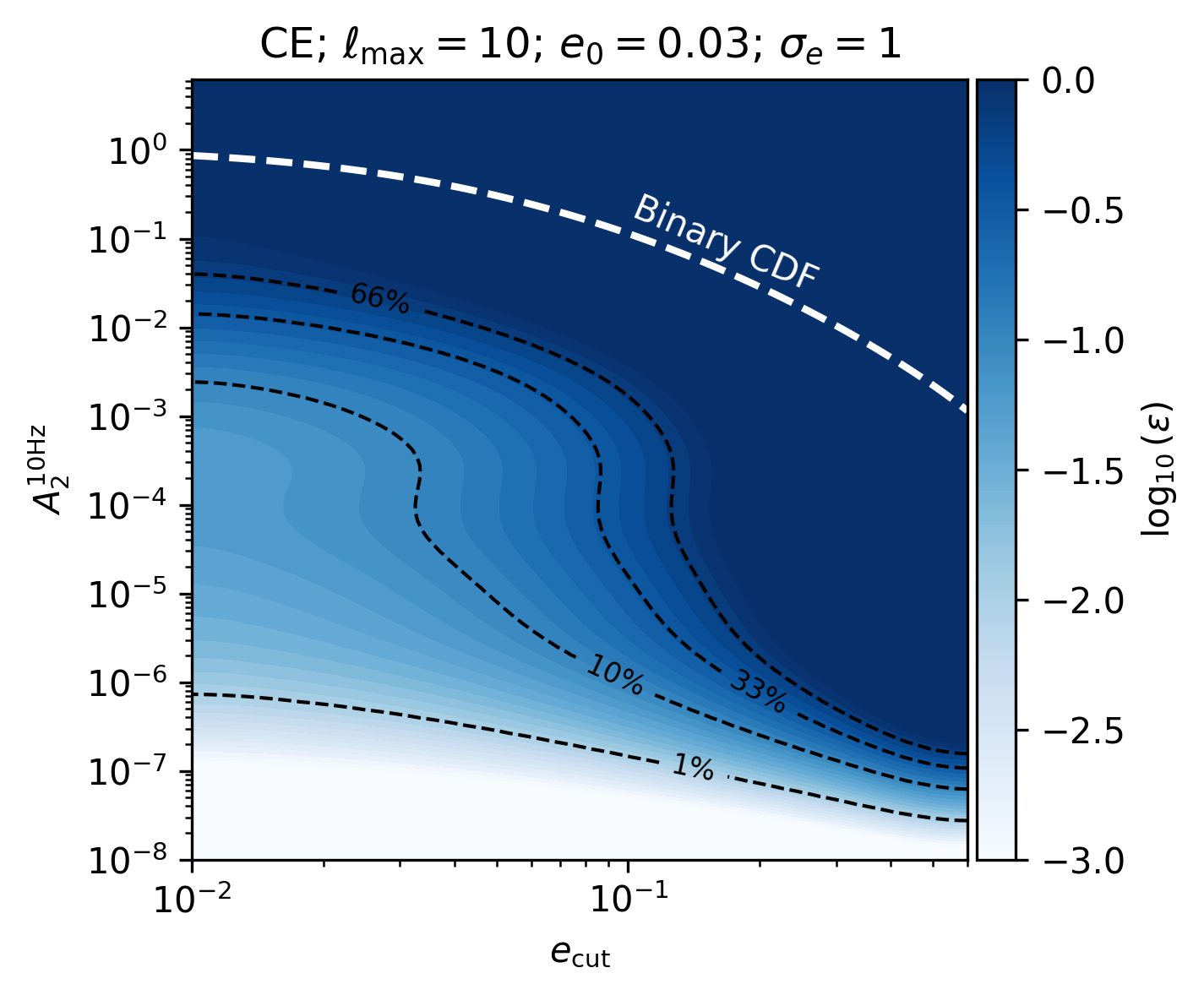}
    \caption{Contours for the fraction $\epsilon$ of GW signals within the tail of a realistic eccentricity distribution (see text), that have a phase shift with $\delta$SNR>3, here for an example dephasing with $n=-13/3$. The high eccentricity tail is defined by a cut--off value $e_{\rm cut}$ and the results are computed as a function of the dephasing amplitude $A_2^{10 \rm Hz}$, here for a representative LVK and CE/ET source consisting of a 8 M$_{\odot}$ + 8 M$_{\odot}$ binary placed at $z=0.2$ and $z=3$, respectively. Note how focusing on the high eccentricity sources increases the chances of detecting weaker EEs. The binary cumulative distribution (CDF) as a function of $e_{\rm cut}$ is over-plotted on the contours to highlight the trade-off between the quantity of sources with $e_{10 \rm Hz}> e_{\rm cut}$ and the detectability boost for EE in sources with high eccentricity.}
    \label{fig:ecc_contours}
\end{figure*}
In the previous section, we have demonstrated that the presence of eccentricity significantly boosts the capacity to detect dephasing caused by EEs. However, this advantage must be weighed against the fact that eccentric signals are intrinsically rare.

The eccentricity distributions of binaries in the various flavours of the dynamical channel have been the subject of much work in the last decade. Most massive stars exist in close binaries accompanied by a distant third star \citep{moe2017,2023pauwels}. In such hierarchical triple systems, the outer companion’s gravitational influence can trigger von Zeipel–Kozai–Lidov (ZKL) oscillations, causing large variations in the inner binary’s eccentricity and inclination \citep{zeipel1910,koz62,lid62}. After the inner pair evolves into a BH binary, these ZKL-driven oscillations can accelerate their merger by enhancing GW energy loss during close encounters. Although GW circularizes the orbit before detection by ground-based observatories, population studies suggest that roughly 1–30 \% of these BBH mergers still retain measurable residual eccentricities when the peak GW frequency reaches 10 Hz \citep{antonini2014a,antognini2016}, { where more precise estimates depend on the chosen detector, the exact definition of eccentricity, and source properties such as redshift \citep{2026MNRAS.545f1938D}}. Globular and nuclear clusters are especially relevant for producing systems with measurable orbital eccentricities through few-body interactions or single–single gravitational captures. Current estimates suggest that roughly 5\% of BBH mergers originating from star clusters may exhibit detectable eccentricity \citep{zevin2019,2024dallamico} {assuming a detectability threshold of 0.05 at 10 Hz}. Interestingly, a non-zero contribution of clusters to the overall merger rate is suggested by the recent discoveries of two events with confidently negative effective spin and component spin magnitude close to the expectation $\chi\sim0.1$ from hierarchically formed black holes \citep{2025Abac}.

Several recent studies have investigated the expected eccentricity distributions of binaries forming and evolving within AGN disks, though the available constraints remain strongly model-dependent:
\citet{Tagawa2021} performed post-processing on their AGN-assisted BBH formation model originally presented in \citet{tagawa2020} to characterize the resulting eccentricity distribution. They found that binary–single (BS) interactions dominate the excitation of eccentricity. Assuming isotropic BS encounters, they predicted that 8–30 \% of BBHs entering the LVK band will have $e_{10,\mathrm{Hz}} \geq 0.03$, and 5–17\% will reach $e_{10,\mathrm{Hz}} \geq 0.3$. If, instead, all interactions are confined to the AGN disk plane, the fraction with $e_{10,\mathrm{Hz}} \geq 0.3$ rises dramatically to 10–70\%.
\citet{Rowan2025_trips} extended this line of work to include gas, analyzing the eccentricities of the most tightly bound binaries in two-dimensional hydrodynamical simulations of binary-single scatterings. While these results are limited by dimensionality and gravitational softening lengths orders of magnitude larger than their BH Schwarzschild radii, they suggest that significant eccentricity can be maintained up to the merger phase, supporting the qualitative conclusions of \citet{Tagawa2021}.
\citet{Wang2025_II} performed detailed two-dimensional post-Newtonian binary–single scattering experiments in AGN-like environments, directly computing $e_{10,\mathrm{Hz}}$ for merged binaries. However, their simulations also employ softening lengths several orders of magnitude larger than the separations at which gravitational waves dominate, which can artificially accelerate inspiral. As a result, their eccentricity estimates should be regarded as upper bounds on the true population-level distribution.
Finally, \citet{Dittmann2025_q_uneq} presented high-resolution hydrodynamic simulations of embedded AGN binaries, deriving empirical relations for $\dot{a}/a$ and $\dot{e}/e$ as functions of binary and disk parameters. Although the validity of these relations once the binary becomes significantly hardened\footnote{Here hardened specifically refers to the binary separation relative to their Hill sphere. At sufficiently low separations, the circum-binary disc will start to become indistinguishible from a circumsingle disc of a single embedded object. The nature of the torques here are likely to become more akin to those of SMBH binaries, although this has not yet been simulated.} is uncertain, they provide a framework for evolving an initial BBH population toward merger. A notable result is that retrograde binaries with near-equal mass ratios can experience rapid eccentricity growth, implying that a non-negligible fraction of AGN-assisted mergers could retain high eccentricity if such systems are common. This corroborates, with improved resolution, the findings of earlier smoothed-particle hydrodynamical simulations \cite{rowan2023} and the 2D grid based work of \cite{Calcino2024}. 
Overall, quantitative predictions vary widely depending on the assumed disk geometry, encounter anisotropy, numerical resolution and a myriad of other parameters related to the AGN disc. However, the emerging picture from these studies is that dynamical encounters and gas torques within AGN disks can produce a broad eccentricity distribution, potentially extending to high $e_{10\mathrm{Hz}}$ values. 

To model the distributions of GW source parameters, we adopt a phenomenological approach in which the overall distribution factorizes as:
\begin{align}
\mathcal{P}(z, \mathcal{M}, \mu, e) \sim \mathcal{P}(z)\,\mathcal{P}(\mathcal{M})\,\mathcal{P}(\mu)\,\mathcal{P}(e),
\end{align}
where the assumptions and limitations are discussed in PI. We consider a representative parametrization for $\mathcal{P}(e)$ motivated by population synthesis and dynamical studies discussed above, i.e. a log-normal distribution characterized by a scatter $\sigma_e$ and a peak value $e_{\rm p}$:
\begin{align}
\label{eq:ecc_distr}
    \mathcal{P}(e_{10 \rm Hz}) \propto \frac{1}{e_{10\rm Hz}}\exp\left(-\frac{\log_{10}(e_{10 \rm Hz}/e_{\rm p})^2}{2\sigma_e^2} \right),
\end{align}
{which is intended to model specific sub-populations of binaries originating from various formation channels.} As reference values, we consider a sub-population of binaries with a high eccentricity tail, where approximately 10\% retain $e_{10 \rm Hz}>0.1$. This is accomplished by choosing e.g. $e_{\rm p}=0.03$ and a scatter of one dex ($\sigma_e=1$), roughly matching the distribution for GW captures presented in \cite{Zevin2021ecc}. Moreover, this choice also aligns with plausible estimates of the eccentricity distribution in the AGN channel, according to our literature review. We stress that any single choice of parameters may not describe the overall distribution of binary eccentricities over multiple formation pathways. However, this can be rectified by simply multiplying Eq. \ref{eq:ecc_distr} with an efficiency factor $\mathcal{E}$ and adding multiple contributions. With this caveat in mind, we can now perform numerical calculations of $\delta$SNRs in the manner described in section \ref{sec:Det_increases} while also considering the effect of the rarity of high eccentricity sources. We adopt a value of $\mathcal{E}=1$ for ease of communication. Finally, we note that while in PI we were able to draw $z$, $\mathcal{M}$ and $\mu$ from their corresponding distributions, this required the evaluation of $\mathcal{O}(10^{8})$ waveforms. This is computationally unfeasible when using accurate eccentric waveforms. Instead, we will use appropriate choices for representative sources for different detectors, as detailed in the corresponding sections.

\subsection{Impact for eccentric populations}
Fig. \ref{fig:ecc_contours} shows the impact of the increased detectability of dephasing in the context of a realistic eccentricity distribution for a sub-population of binaries. We show the results for a dephasing prescriptions with $n=-13/3$, which transfer qualitatively to other choices. The four panels show the fraction $\epsilon$ of GW signals within the tail of the eccentricity distribution that showcase a phase shift with significant $\delta$SNR. More precisely, the high eccentricity tail is defined by introducing a cut--off value $e_{\rm cut}$, and $\epsilon$ represents the fraction of sources with eccentricity above the cut-off that additionally have $\delta$SNR > 3. The results are computed as a function of the dephasing amplitude $A_2^{10 \rm Hz}$, for a representative LVK and CE/ET source consisting of a 8 M$_{\odot}$ + 8 M$_{\odot}$ binary placed at $z=0.2$ and $z=3$, respectively. Changing these choices does not modify the results beyond slightly shifting the contours vertically due to the different overall SNR of the source.



The contour plots demonstrate the significant advantages of searching for EE in the high eccentricity tail of the distribution of GW signals. As an example, we now focus on the top left panel, representing the results for LVK A+ sensitivity. When considering the majority of sources (i.e. choosing a small value for $e_{\rm cut}$) we can expect to have significant EEs in the majority of signals only if the corresponding dephasing reaches a magnitude of $A_2^{10 \rm Hz}\sim1$. However, when considering sources above a moderate eccentricity threshold $e_{\rm cut} >0.2$, the required dephasing magnitude reduces to few $10^{-2}$. This reduction corresponds to a significant increase in the phase space volume of physical parameters that result in a detectable EE and the consequences of this fact will be discussed in section \ref{sec:conclusions}. More drastic results are seen for the other detector configurations of LVK A\#, ET and CE. In particular, the latter shows the largest increase in the detectability of EEs. For CE, the value of $A_2^{10 \rm Hz}$ required to reach $\delta$SNR>3 in the majority of signals drops by four order of magnitudes when considering sources above the eccentricity threshold of $e_{\rm cut} =0.2$. Additionally, we stress that these results are computed for $\ell_{\rm max}=10$, which is the typical maximum harmonic modeled in currently available waveform templates. The results scale extremely strongly with the maximum harmonic, as shown in section \ref{sec:Det_increases}.

Fig. \ref{fig:ecc_contours} highlights an interesting trade-off: 1) The number of sources above a certain eccentricity cut at 10 Hz, and 2) the number of sources that are affected by EE capable of producing dephasing with a certain magnitude. As an example, approximately 3\% of binaries retain an eccentricity at 10 Hz above 0.2 for our choice of distribution parameters. However, such binaries would showcase significant dephasing even when perturbed by an EE that is $\mathcal{O}[10^2]$ ($\mathcal{O}[10^4]$) smaller than for the corresponding circular signal in LVK (CE/ET), greatly increasing the plausibility of detection. We propose that this balance  would uniquely reflect the physics of different binary merger channels, and therefore provide a framework to connect astrophysical source populations with measurable waveform features. We will further discuss the consequences of this aspect in section \ref{sec:conclusions}.

\section{Discussion and conclusion}
\label{sec:conclusions}
\subsection{Implications of an increased detectability of EEs}
The enhanced sensitivity to dephasing in eccentric signals implies that EEs can become relevant across a broader region of parameter space. In PI, we identified threshold values of the environmental parameters $\xi_i$ (see Table~\ref{tab:placeholder}) at which a significant fraction of circular sources exhibit detectable EE signatures, depending on the detector configuration. Here, we recompute these thresholds for an eccentric signal. As a representative case, we consider a binary retaining an eccentricity of $e_{10{\rm Hz}} = 0.3$, consistent with the eccentric candidate events GW190701, GW200129, and GW200208 \citep{2024gupte}. We focus on Roemer delays (line-of-sight accelerations), which are pertinent to both dynamical and AGN-assisted formation channels. Our analysis shows that for a dephasing prescription with $n=-13/3$, such eccentricity yields a dephasing amplification factor $\delta{\rm SNR}$ of order a few $\times 10^{2}$ for LVK and up to $\sim 10^{5}$ for ET/CE, assuming $\ell_{\rm max}=10$.

Table 2 of PI lists the critical $\xi_i$ values required for circular binaries to exhibit measurable dephasing. For Roemer delays, these correspond to $\xi_{\rm R} \sim 10^{7}$ M$_{\odot}$ AU$^{-2}$ for LVK and $\xi_{\rm R} \sim 10^{2}$ M$_{\odot}$ AU$^{-2}$ for CE/ET. While the former is implausible for standard dynamical formation channels, the latter is compatible with, for instance, a tertiary $10$ M$_\odot$ black hole at a separation of $\sim 0.3$ AU, well within the characteristic range $R \sim 10^{-2}$–$10^{3}$ AU typical of stellar clusters \citep{antonini2016}. An increase in detectability by $\sim 10^{5}$ extends the accessible tertiary separation by a factor of $\sim 300$, to approximately $10^2$ AU. Given that the tertiary separations are log-normally distributed within the characteristic range \citep{trani2022}, this range encompasses {a significant fraction of} physically plausible parameter space, {and also reaches the scales of field hierarchical triples \citep{antonini2017}}. In other words, for moderately eccentric signals, EEs become an expected signature of the stellar cluster and hierarchical triple formation channel for next generation ground based detectors. These considerations apply even more strongly to dynamical sub-channels that require closer encounters, such as GW captures and chaotic scatterings \citep{kai2024}.

The thresholds for $\xi_{\rm R}$ can also be applied to the AGN channel. Consider a massive black hole (BH) of mass $M_{\bullet}$ hosting a binary at a separation of $N_{\bullet}$ Schwarzschild radii. The corresponding parameter is:
\begin{align}
    \xi_{\rm R} \approx 2.6 \times 10^{8} \left(\frac{10^7 \, \rm{M}_{\odot}}{M_{\bullet}} \right) \frac{1}{N_{\bullet}^2} \, \left[{\rm{M}_{\odot}}  {\rm AU}^{-2} \right].
\end{align}
For LVK A+ and A\# sensitivities, such accelerations would only be detectable for binaries orbiting within a few Schwarzschild radii\footnote{This may offer the most plausible interpretation for the acceleration reported in \citet{2024Han} for GW190814, although we note that the underlying data analysis presents several methodological issues (private communication; details will be published in forthcoming work).} of the event horizon of typical massive BHs \citep[see e.g.][for constraints on the massive BH mass function]{Greene:2007dz}.
However, the $\delta$SNR enhancement provided by a moderately eccentric signal extends this detectable range to separations of several hundred Schwarzschild radii. This scale is particularly relevant, as it coincides with the predicted location of the inner migration trap in most AGN disc models \citep{sirko2003,bellovary2016,mckernan2018,tagawa2020}, where the efficiency of mergers is expected to be the greatest \citep{Fabj24}. For next-generation detectors, which could be sensitive to Roemer delays as small as $\xi_{\rm R} \sim 10^{-2}$ M$_{\odot}$ ${\rm AU}^{-2}$ for moderately eccentric binaries, dephasing remains significant out to $N_{\bullet} \sim 10^{5}$. This is comparable to the scale at which AGN discs are expected to become self-gravitating \citep{Shakura:1973uy,sirko2003,2007levin}. Consequently, for CE/ET-class observatories, dephasing from Roemer delays should represent an expected signature of the AGN channel in moderately eccentric binaries, irrespective of whether the binary originated in a migration trap or formed via in situ star formation near the self-gravity radius \citep[see e.g.,][]{2023derdzinski}. Similar considerations apply to the various gas-induced EEs discussed in PI, which constitute distinctive signatures of the AGN channel. Notably, eccentric signals substantially widen the region of AGN disc parameter space in which gas interactions yield detectable imprints, as illustrated by the comparison with the results presented in Fig.~5 of PI with respect to this work. Here we refrain from an extensive discussion, as AGN disk models depend on a wide range of parameters, and simply note that our qualitative conclusions hold regardless. In particular, an increase in the detectability of dephasing translates directly into a linear reduction in the gas density required to produce a measurable effect.

\subsection{Suggestions for the inference of EEs in current and upcoming GW catalogues}
Our findings can be used to inform both current and future parameter inference studies. Starting with next-generation detectors such as ET and CE, our analysis demonstrates that the presence of moderate eccentricity in a signal dramatically enhances the detectability of EEs, to a point where they become an expected observable for binaries formed in the dynamical and AGN channels. In the former, the majority of the plausible tertiary separation range yields detectable signatures for eccentricities $e_{\rm 10Hz} \gtrsim 0.2$. In the latter, dephasing due to Roemer delays becomes observable throughout the entire disc, and even gas-induced effects are expected to produce measurable imprints across a significantly broader parameter space of densities and temperatures. While sources above $e_{\rm 10Hz} \gtrsim 0.2$ may only represent a small fraction of the entire detectable binary populations, CE/ET are expected to detect hundreds of thousands of events. Therefore, we can plausibly expect significant dephasing due to a wide range of EEs in hundreds of signals, and even more if advances in eccentric waveform modeling can extend the range of $\ell_{\rm max}$.

Our conclusions are less drastic for current and near-future LVK observing runs, though they highlight a specific opportunity to verify the origin of already identified candidate eccentric signals. These are constituted by a handful of signals within the $\mathcal{O}(10^2)$ events confirmed up to O4a.
For current catalogs, we argue that the immediate priority should be to robustly confirm eccentricity in candidate systems, followed by targeted searches for dephasing signatures due to Roemer delays in the higher harmonics of their gravitational-wave emission. While eccentricity constitutes a clear signature of dynamical formation, by itself it is not sufficient to distinguish among sub-channels nor between so called "dry" formation or the AGN pathway. Instead, this additional information can be extracted from a detection of EEs. We provided a practical estimate of the potential $\delta$SNR enhancement in signals with $e_{\rm 10Hz} \gtrsim 0.2$, which is obtained by considering the maximum contributing harmonic number. For Roemer delays, the $\delta$SNR enhancement factor scales as $\ell_{\rm max}^{16/3}$, up to when the power of each harmonic is saturated. Because of this, we argue that a joint detection of eccentricity and line-of-sight acceleration is already plausible within current GW catalogs, in the case that the AGN channel contributes a significant fraction of compact-object mergers, and that those mergers take place in the inner migration trap of AGN disks. This configuration would produce a handful of events with moderate eccentricity (see section \ref{sec:Det_increases_pop}) and with $\xi_{\rm R} \gtrsim 10^4$ M$_{\odot}$ AU$^{-2}$ (see previous section). In principle, this subset of sources could constitute the entirety of current eccentric candidates. If confirmed, such a measurement would represent the most compelling evidence for an AGN origin of gravitational-wave sources observed so far.

\section*{Acknowledgments}
L.Z. is supported by the European Union’s Horizon 2024 research and innovation program under the Marie
Sklodowska-Curie grant agreement No. 101208914. K.H, P.S., and J.S. are supported by the Villum Fonden grant No. 29466, and by the ERC Starting Grant no. 101043143 – BlackHoleMergs. J.T. acknowledges support from the Horizon Europe research and innovation programs under the Marie Sk\l{}odowska-Curie grant agreement no. 101203883. J.T. was supported by the Alexander von Humboldt Foundation under the project no. 1240213 - HFST-P. The Center of Gravity is a Center of Excellence funded by the Danish National Research Foundation under grant No. 184. The research leading to this work was supported by the Independent Research Fund Denmark via grant ID 10.46540/3103-00205B.

\bibliographystyle{aasjournal}
\bibliography{main}

\end{document}